\documentclass{aastex}

\usepackage{psfig}

\newcommand{\bc}{\begin{center}}
\newcommand{\ec}{\end{center}}
\newcommand{\bea}{\begin{eqnarray}}
\newcommand{\eea}{\end{eqnarray}}

\def\beq{\begin{equation}}
\def\eeq{\end{equation}}

\def\ben{\begin{enumerate}}
\def\een{\end{enumerate}}
\def\bit{\begin{itemize}}
\def\eit{\end{itemize}}

\def\H0{H_0}
\def\om0{\Omega_{m,0}}
\def\omb0{\Omega_{b,0}}

\def\oml0{\Omega_{\rm \Lambda,0}}

\def\U{U_{300}}

\newcommand{\pz}{{\it pixel-z} }

\newcommand{\dlz}{d_{\rm L}(z)}

\newcommand{\pd}{{\rm d}}

\def\dfrac#1#2{{\displaystyle\frac{#1}{#2}}}

\newcommand{\Msun}{M_{\sun}}

\bibpunct[,]{(}{)}{;}{a}{}{,}

\begin{document}

\title{The Star Formation History of Galaxies Measured from Individual
Pixels.\\ I. The Hubble Deep Field North}


\author{Alberto Conti${^1}$, Andrew J. Connolly, Andrew M. Hopkins${^2}$}
\affil{Department of Physics and Astronomy, University of Pittsburgh,\\
3941 O'Hara Street, Pittsburgh, PA 15260}

\author{Tamas Budav{\'a}ri, Alex S. Szalay, Istv{\`a}n Csabai}
\affil{Department of Physics and Astronomy, John Hopkins University,\\
3400 North Charles Street, Baltimore, MD, 21218}

\author{Samuel J. Schmidt, Carla Adams${^3}$, Nada Petrovic${^4}$}
\affil{Department of Physics and Astronomy, University of Pittsburgh,\\
3941 O'Hara Street, Pittsburgh, PA 15260} 

\altaffiltext{1}{Currently at Computer Sciences Corporation, Space Telescope Science Institute,
  3700 San Martin Drive, Baltimore, MD 21218}
\altaffiltext{2}{Hubble Fellow}
\altaffiltext{3}{Currently at the New Mexico State University, MSC 4500
Astronomy P.O. Box 30001, Las Cruces, NM  88003}
\altaffiltext{4}{Currently at the Department of Astronomy and Astrophysics,
5640 S. Ellis Ave, Chicago, IL 60637}



\shorttitle{Pixel-z: The Hubble Deep Field North}
\shortauthors{Conti at al.}
\singlespace

\begin{abstract}
We analyze the photometric information contained in individual pixels of
galaxies in the Hubble Deep Field North (HDFN) using a new technique,
{\it pixel-z}, that combines predictions of evolutionary synthesis models
with photometric redshift template fitting. Each spectral energy
distribution  template is a result of modeling of the detailed physical
processes affecting gas properties and star formation efficiency. The
criteria chosen to generate the SED templates is that of sampling a
wide range of physical characteristics such as age, star formation rate,
obscuration and metallicity. A key feature of our method is the
sophisticated use of error analysis to generate error maps that define
the reliability of the template fitting on pixel scales and allow for
the separation of the interplay among dust, metallicity and
star formation histories. This technique offers a number of advantages over
traditional integrated color studies. As a first application, we derive
the star formation and metallicity histories of galaxies in the HDFN.
Our results show that the comoving density of star
formation rate, determined from the UV luminosity density of sources in
the HDFN, increases monotonically with redshift out to at least redshift
of 5. This behavior can plausibly be explained by a smooth
increase of the UV luminosity density with redshift coupled with an
increase in the number of star forming regions as a function of redshift.
We also find that the information contained in individual pixels
in a galaxy can be linked to its morphological history. Finally, we
derive the metal enrichment rate history of the universe and find it in
good agreement with predictions based on the evolving HI 
content of Lyman-$\alpha$ QSO absorption line systems.
\end{abstract}

\keywords{cosmology: observations --- galaxies: distances and redshifts ---
galaxies: evolution --- galaxies: formation --- galaxies: luminosity
function}

\section{Introduction}
In recent years the photometric redshift technique has become a viable
and cheap alternative to spectroscopic redshifts
\citep{loh86,con95a, lan96, gyw96, mob96, saw96, bal00, fur00, bud01,
fer01, ric01, leb02}. 
In all those cases where galaxies are too faint for spectroscopic studies, 
the photometric redshift technique remains the only practical way of
estimating galaxy redshifts. With ongoing large multicolor
observational programs such as the Sloan Digital Sky Survey and 2MASS,
the photometric 
redshift technique will prove to be an even more powerful tool in
interpreting these increasingly large and detailed data sets. The use of
photometric redshifts will open the door to the study of multivariate
distributions (such as the galaxy luminosity function) and may in fact
provide statistical redshift distributions for any population of
galaxies in any environment \citep[ in preparation]{sch02}.

Here we focus on what is perceived by many as a byproduct of
the photometric redshift technique, namely the best fitting spectral
energy distribution (SED) template. This work takes as its starting
point a new approach in the use of the photometric redshift technique
that extends and expands its original purpose. Our motivation resides in 
recent technical and conceptual developments that make this new step
possible. On one hand, the use of sophisticated training
algorithms \citep{bud00} allows for a
better interpretation 
of the intrinsic photometric properties of a sample. This step shifts
the focus from the photometric redshift technique in itself to the
actual properties of the sample. Secondly, deep surveys such as
the Hubble Deep Fields\footnote{Based on observations with the
NASA/ESA \emph{Hubble Space Telescope} obtained at the Space Telescope
Science Institute, which is operated by the Association of
Universities for Research in Astronomy, Inc., under NASA contract NAS
5-26555.} \citep{wil96,wil00} have stimulated the
growth of ideas, new techniques and ``proof of concept'' papers. One of
these papers \cite[ hereinafter ABR99]{abr99}, focused on ``{\it exploring the
resolved multicolor data for galaxies of known redshift using
spectral-synthesis models}'' thereby dramatically shifting the focus of
the photometric redshift technique to the reliability of the template
spectra used in the fitting.

Our work takes ABR99 as a starting point. However, instead of focusing
as ABR99 did on the understanding of the relative ages of bulges and disks,
and the formation history of elliptical galaxies via spatially resolved
colors, we focus onto two direct enhancements of the ABR99 technique and
two distinct applications:
\bit

\item[$\vartriangleright$]
Use of all available redshifts: ABR99's sample was, at the time, the
faintest statistically-complete spectroscopic redshift sample currently
available in the HDFN. We extend his sample by including not only the
$\sim 190$ object in the HDFN with measured spectroscopic redshifts, but
all of the $\sim 1500$ galaxies detected in the HDFN for which we will
measure photometric redshifts. While this approach propagates the larger
redshift uncertainties onto the measured properties of galaxies, the use
of the original HDFN passbands extended by the NICMOS data reduces the
redshift uncertainty to more than acceptable levels $\sigma_z \sim
0.05$ \citep{con97}. 
Furthermore, as we will show in Section~\ref{sec:analysis}, Monte
Carlo realizations of the photometric redshift uncertainty on all galaxies
in the HDFN with measured spectroscopic redshift, show that the
derived properties of galaxies are not adversely affected by
uncertainties in redshift as large as $\sigma_z=0.2$.

\item[$\vartriangleright$]
Wide template range: By extending the 4-filter color approach of ABR99
to a 6-filter 2160 template approach, we allow for a more detailed
description of the individual characteristics of the object being
compared to the set of templates. In particular our description will
allow for variations in age, dust content, metallicity and star
formation rate on a wide range of scales. As a consequence of this
``redundant'' approach, we will be able to resolve subtle
differences within an individual galaxy's stellar populations.
This is one of the goals of our approach. In addition, the large number
of templates used will allow for a multivariate error analysis on a
pixel-by-pixel level. This in turn will produce detailed error maps for
each HDFN galaxy. We will make use of such maps to assess the accuracy
and the significance of the physical characteristics derived from each
pixel's SED.

\item[$\blacktriangleright$]
 Science from individual pixels: Each pixel can be regarded as the smallest
``multicolor unit'' available in the HDFN. We will make use of the multicolor
information for each of the pixels in a galaxy to constrain the relative
ages for physically distinct sub-components of the galaxy and determine
their dust and metallicity content. As a byproduct, in
Section~\ref{sec:method} we will also be able to determine which
pixel, and hence which parts of a galaxy, are contributing to the
estimate of the photometric redshift. This feature has the potential of
being able to separate projected objects that are close only in angular
distance.

\item[$\blacktriangleright$]
Comoving density of star formation and metals: The comoving density of star
formation and metallicity enrichment are calculated as a function of 
redshift using individual pixels in the HDFN. We will compare the
results obtained by other groups using standard techniques with our own
approach. In addition, we will compute the luminosity function of
galaxies in redshift intervals, integrate it out to faint magnitudes and
recover the luminosity function correction to be applied to our derived
star formation rates (SFR) to compensate for the way different magnitude
ranges are sampled at different redshifts. 
\eit 

The outline of the paper is as follows. In Section~\ref{sec:method} we
will illustrate the \pz method and why we have decided for its first
application on the Hubble Deep Field North. In Section~\ref{sec:temp} we
describe the strategy and trade offs necessary to create a set of template
spectral energy distributions that cover a wide range of physical
characteristics. Section~\ref{sec:analysis}
presents the results of the \pz method including the maximum likelihood
error analysis we used to disentangle uncertainties and
degeneracies introduced by the use of a large set of templates. In this
section we will also compute, based on our findings, the star formation
rate of all pixels in the HDFN. In Section~\ref{sec:madau} we calculate
the star formation history of all pixels in the HDFN. Furthermore, we
derive the luminosity function in redshift bins to account for the
large fraction of missing luminosity at high redshift. Finally, we
calculate the comoving metal enrichment rate for all pixels in the HDFN.
Section~\ref{sec:end} presents the conclusions.

Throughout the manuscript, we assume a matter dominated universe with
$\Omega_m=1$ and $H_0 = 75\ h$ km s$^{-1}$ Mpc$^{-3}$.

\section{The Pixel-z Method}
\label{sec:method}
By making use of strong spectral features such as the $4000$\,\AA\ break,
the Balmer break and the Lyman decrement, the standard template-fitting
photometric redshift techniques is able to return a fast and accurate
estimate of a galaxy's redshift. Generally, the fitting function
takes the following form:
\begin{equation}
\label{eq:lkhood}
\chi^2(z,T) = \sum_{i=1}^{N_{\rm f}}
\frac{[F_{{\rm obs},i} - b_j \times F_{i,j}(z,T)]^2}{\sigma^2_i}
\end{equation} 
where $F_{{\rm obs},i}$ is the flux through the ith filter, $b_j$ is a
scaling factor, $F_{i,j}$ is the flux through the ith filter
of the jth spectral energy distribution template (calculated at a redshift
$z$) and $\sigma_i$ is the uncertainty in the observed flux. The sum is
carried out over all available filters $N_{\rm f}$.
The resulting $\chi^2$ is then minimized as a function of
template $T$ and redshift $z$ providing an estimate of the redshift of the
galaxy and its spectral type (together with the variance on these
measures). 

The accuracy of the photometric redshift estimate obtained by applying
equation (\ref{eq:lkhood}) to galaxy data, critically depends on the
number and wavelength coverage of the passband filters used, and on the
signal-to-noise characteristics of the source.  At low redshifts ($z
\lesssim 0.7$) it is important to cover the ultra-violet and optical
region of the spectrum in order to be able to ``bracket'' the
$4000$\,\AA\ break with the U, B and V passbands. At very high redshifts
($2.5<z<5.0$) the ultra-violet region becomes very important as the
$912$\,\AA\ break is shifted into the $\U$ passband filter
\citep{mad96,ste96}.  At intermediate redshifts adequate coverage of the
spectral features is guaranteed by infrared filters: the J filter being
instrumental for galaxies at $z \lesssim 1.5$, while the H and K filters
extend the redshift coverage past $z \sim 2.5$. Provided UV and IR filters
are available, at least four filters are needed to identify the strong
spectral features required to obtain a statistically significant
estimate of the redshift. However, modern deep surveys can
boost this number to as many as six, thereby significantly improving the
reliability of the redshift estimates at any redshift \citep{hog98}.

Generally, the observed flux $F_{{\rm obs},i}$ is taken to be that of the
entire galaxy in each bandpass obtained by adding the contribution of
all pixels (identified as being part of the galaxy) above some chosen
threshold\footnote{Typically an arbitrary 3$\sigma$ cut above sky level
is chosen, but see also \citet{hop01} for a more rigorous definition of
detection threshold.}. 
Commonly used object detection software
packages, in fact, not only identify and catalog objects, but also
produce a list of pixels belonging to each of the identified
objects. This catalog of pixels for each galaxy will be the starting
point of our analysis. 

The information required to estimate the
photometric redshift of an object does not make direct use of the pixel
information and is limited to 
the objects' observed magnitudes and relative measurement
errors. However, the flux of individual pixels
could, in principle, be treated as a separate ``source'' within the data
and, thanks to the quite general form of the fitting function given by 
equation (\ref{eq:lkhood}), a photometric redshift for individual
pixels can be computed.\footnote{We should mention that the analysis of
galaxy images pixel by pixel has the potential disadvantage of introducing 
larger photometric uncertainties than total fluxes measured in apertures.
Ideally one would like to keep the spatial information present in the
image along with a better signal-to-noise ratio. A possible solution is
described in \citet{bud02}, where spatially connected pixels of similar
colors are joined into {\em{}superpixels} in order to improve on the
statistical errors without mixing the different galaxy components, e.g.\
the red bulge or the bluer star-forming regions in spiral arms.}

While the extension of the photometric redshift
technique to individual pixels is straightforward (modulo the photometric error
propagation), the real issue resides in the
interpretation of the results. Can the use of photometric redshifts
computed on individual pixels provide new insights into the intrinsic
properties of galaxies? 
Moreover, once a photometric redshift for a pixel inside a galaxy has
been determined, what is its significance in terms of the global
photometric or spectroscopic redshift of the entire galaxy? Finally, how
does the photometric redshift error propagate on a pixel scale?

This work addresses all the aforementioned questions. In particular, we
have developed a new technique called \pz to take advantage of the
information contained in individual pixels that 
\begin{enumerate}
\item enables the deblending of systems associated by angular
superposition alone;
\item enables one to decompose the internal photometric 
structure of observed galaxies into basic constituents such as the age
of the stellar population, their metallicities and their dust content;
\item enables, under simplifying assumptions, the determination of the
star formation rate for individual pixels inside a galaxy;
\item returns the contribution of each pixel to the star
formation history of a galaxy thereby allowing, for a sufficiently
large sample of galaxies, a direct assessment of the drivers behind the
current understanding of the global star formation history plot.
\end{enumerate}

For the first application of this technique, we choose to use the
Hubble Deep Field North (HDFN). This rich dataset has many of the
characteristics that make it an ideal testbed for our purposes. First and
foremost, it is among the best studied areas of the sky and, as such, can
be regarded an ideal benchmark for our technique. Moreover,
the HDFN provides not only one of the deepest images for multi-color
selection of ``Lyman break'' galaxies but also morphological and size
information for high redshift galaxies. For our analysis we will make
use of the 2500$\times$2500 pixel versions of the HDFN and NICMOS data
together with the relative weight maps.
The latter are needed to correctly estimate the photometric errors on
individual pixels. We use version 1.04 of the renormalized rms maps.
These maps give the ``true'' noise level in the data,
corrected for inter-pixel correlations. These correlations, induced
both by drizzling and by the use of convolution kernels during the
reduction process, reduce the apparent pixel-to-pixel noise level in the
images, making it difficult to assess flux uncertainties without knowing
the correction factors.
With the rms images, one can sum (in quadrature) the noise values for
pixels over any aperture and determine the effective flux uncertainty
within that aperture or use the measure on a pixel-by-pixel basis. The rms maps
quantify the uncertainty due to noise in the background. These maps are
used to determine the detection significance of an object or a pixel.
We shall return to this point later on when we will show how the error
in the pixels of an object will allow for the construction of an
object's error map used to assess the significance of our photometric
decomposition. 

The results of the application of the photometric redshift technique to
individual pixels in galaxies in the HDFN is shown in 
Figure~\ref{fig:phist1} and Figure~\ref{fig:phist2}. 
For each pixel a photometric redshift is
computed using the template fitting technique. The number of templates
used in this particular case is rather small, being the four empirical SEDs
compiled by \citet{col80} (hereafter CWW). However, as we shall see
later on, this does not change the reliability of this particular part
of our analysis. For each of the pixels,
the resulting photometric redshift estimate is simply a function of the
flux in each of the available band-passes. When applied to the entire
galaxy (i.e. the sum of the contributions from individual pixels), the
photometric redshift returned should correlate with the number of pixels
that have photometric redshifts near the galaxy overall redshift. For
the galaxy in Figure~\ref{fig:phist1}, the measured {\it spectroscopic}
redshift is $z=1.013$, while the measured {\it photometric} redshift is
$z=1.061 \pm 0.05$. The galaxy shown is rather bright ($F160W \sim
19.88$), and most of the pixels in the galaxy appear tightly distributed 
around $z\sim 1$, peaking at a redshift just greater than one with a
rather tight $FWHM=0.11$. This value of the FWHM is typical of galaxies
in the HDFN. In fact, almost all galaxies in the HDFN display this tight
unimodal distribution. The entire sample is characterize by a typical
$FWHM = 0.16$.

While this behavior was somehow expected by ``construction'', perhaps the
more interesting features reside in the two other peaks at higher redshift
and the information they might contain. One can envision a scenario in
which two galaxies at different redshifts, but close in angular distance,
might be separated by using this method. We will address this issue in a
forthcoming paper. In this particular case, the two ``extra'' peaks do
not belong to separate objects but are a direct result of the lower
signal-to-noise ratio of the outskirts of the galaxy, which in turn are
best fitted by SEDs at much higher redshift.

Figure~\ref{fig:phist2} on the other hand, displays a different
behavior. This galaxy appears in the HDFN images as either two or
possibly three separate objects. The distribution of pixels reflects
this multiple object hypothesis. Three peaks of comparable height are
clearly recognizable at redshift $z\sim 0$, $z\sim 0.85$, $z\sim 3$. The
template distribution shows that either we are in the presence of a
single object at spectroscopic redshift $z=0.130$ \citep{dic02} with
distinct and spatially separated stellar populations, or we are in the
presence of separate objects that have been erroneously identified as a
single galaxy. The latter hypothesis is strongly supported by direct
comparison with the work of \citet{con99}. In fact, the component at
$z\sim 0$ has been identified by \citet{con99} as the most likely low
redshift quasar candidate in the HDFN (ID0094). This
suggests that we are in the presence of at least 2 objects, one of which
has been correctly identified via template fitting as ``stellar-like''. 

These simple examples portray the extended reach of this technique.
Our present work, however, takes Figures~\ref{fig:phist1} and
\ref{fig:phist2} as our starting point and 
focuses on the {\it relative} properties of each galaxy by fixing the
redshift of all pixels to be that of the whole galaxy, determined either
photometrically or spectroscopically.
This approach is equivalent to removing the redshift dependence in
equation (\ref{eq:lkhood}) and solving for the best fitting template
type. Removing the redshift dependence, however, {\it does not}
eliminate the existing covariance between redshift and SED template. 

Generally, the resulting best fitting template is treated as a
byproduct of the photometric redshift estimate, the redshift being the
focus of the fitting. This is in part due to the prevalent small number of
templates used (typically local galaxy data and hence not
entirely representative of the large variety of SEDs observed at
different epochs), and in part to the use of evolutionary synthesis
models which remain untested outside the standard Hubble types.

The current implementation of the \pz technique, shifts the attention to
the SED templates. In fact, the SED of a galaxy should reflect the
distribution of stellar masses, ages and metallicities and hence provide
clues to the past history of star formation. Building on this fact, we
will show in the next Sections that, by fitting SEDs to individual
pixels in a galaxy, we can recover the morphological
characteristics of the galaxy and, perhaps more importantly, separate
the individual contributions of age, metallicity, dust and star formation
history.

\section{Template Construction}
\label{sec:temp}
We use the most recent \citet{bru00} models to generate a large number
of SEDs that will be used for template fitting for each of the
pixels of all the HDFN galaxies. Each SED is a result of modeling of the 
detailed physical processes affecting gas properties and star
formation efficiency. The modeling is generally based on assumptions about
the stellar birthrate. The adjustable parameters in the models are typically
the initial stellar mass function (IMF), the star formation rate and
chemical enrichment.

The criteria we have chosen to generate the SED templates is that of
maximizing our ability to solve for the aforementioned
quantities. Specifically, we generate SEDs with the following
characteristics: 
\begin{enumerate}
\item we allow the underlying stellar population within each pixel
to vary over a wide age range. Typically from extremely young
($0.001, 0.01, 0.1, 0.5$ Gyrs), to middle age ($1,3,5$
Gyrs), to old and very old ($9, 12, 15$ Gyrs), for a total of ten 
ages;
\item we assume that the fluxes of individual pixels can be
modeled using an exponential star-formation rate with an e-folding
timescale $\tau$, i.e. $\Psi(t) = \Psi_0 e^{(-t/\tau)}$. This
parametrization turns out to be quite convenient for its simplicity in
describing the star formation rate of an instantaneous burst when $\tau\to
0$ and a constant star formation when $\tau\to\infty$. The e-folding
times used for $\tau$ are $0.1$ Gyrs for an extremely short bursts,
$1,3,5,9$ and $12$ Gyrs for subsequently longer bursts.
Is is worth noting at this point that an exponential star formation
rate for individual pixels does not inevitably lead to an exponentially
decaying star formation rate for the galaxy as a whole other than in the
special case in which every pixel in the galaxy is coeval and all have
the same star-formation rate;
\item since pixels with any star formation history can be expanded in
series of instantaneous bursts, each having fixed metallicity,
the spectral evolution of individual pixels (or whole galaxies) can be
investigated without prior knowledge of chemical evolution. We
assume the SEDs to be characterized by six possible metallicities,
ranging from $\frac{1}{50}$ to $2.5$ times that of the sun;
\item the general spectral characteristics of the SEDs of galaxies will
be modified by the presence of dust. We parametrize dust obscuration
in terms of the relative optical extinction in the rest frame $E_{B-V}$
using the reddening curve $k(\lambda) = A(\lambda)/E_{B-V}$ for
star-forming systems formulated by \citet{cal00}. For each of the SEDs
we allow for six independent values of extinction ranging from no
extinction to $0.9$ magnitudes of extinction.
\end{enumerate}
In all our templates, we assumed a Salpeter IMF with low and high mass
cut-offs of 0.1 $M_{\odot}$ and 125 $M_{\odot}$ respectively.

\section{Analysis}
\label{sec:analysis}
The majority of the galaxies in the HDFN do not have measured
spectroscopic redshifts. These, in fact, amount only to $\sim 190$, while
the total number of galaxies cataloged is $\sim 1500$. Therefore, we make use
of equation~(\ref{eq:lkhood}) together with the CWW templates to determine
the redshift of all galaxies in the HDFN. 
We then generate SEDs that span 10 ages, 6 star formation
e-folding times, 6 metallicities and 6 color extinctions for a total of
2160 templates. We fit each of the 2160 SED templates to all the
pixels in each of the HDFN galaxies maintaining the redshift of all
pixels {\it fixed} to that of their host galaxy during the entire
procedure. This latter step effectively removes one of the degrees of
freedom in our fit and returns the properties of each of
the pixels in terms of their best fitting template.

The best fitting template is
calculated using a least squares maximum likelihood estimator similar in
spirit to the one of equation~(\ref{eq:lkhood}), but with a much larger
number of SED templates, specifically:
\begin{equation}
\label{eq:lkhood2}
{\cal L}({\cal T}) = 
\prod_{n=1}^{N_{\rm f}} {1\over \sqrt{2\pi}\sigma_i}
 \exp\left\{ - {[F_{{\rm obs},i} - b_j \times F_{i,j}(z_{gal},{\cal T})]^2
\over 2 \sigma_i\,^2}\right\}
\end{equation} 
where the SED templates ${\cal T} = {\cal T}\{t,\tau,E_{B-V},[Fe/H]\}$ now
explicitly accounts for the dependence on age, e-folding time,
extinctions and metallicity. The redshift of all pixels $z_{gal}$ is
held fixed. Before we can interpret the results of this fit, however, 
we need to state clearly the assumptions and simplifications 
we make while computing the fit. 

Firstly, we make the simplifying assumption that the all pixels in the
galaxy are coeval, that is, we explicitly assume that the galaxy as
a whole, and hence each pixel within it, has a common age.
This assumption is not required by our analysis and
is not a limitation of our approach. Instead it serves as a
simplifying starting point for the first application of the \pz
technique. The actual ``ages" being fit to in each pixel in a galaxy
are a result of the luminosity-weighted ages of the stellar
populations probed within the pixel, that produce the observed pixel colors.
The ``age" referred to by the synthesis model used, though, is simply
the time since the onset of initial star formation in the model. The
relationship for a passively-evolving system is straightforward, but
for a galaxy where knots of recent star-formation are visible, or
which has experienced a complex star formation history, the relationship
is much less clear. The assumption of a common age for all pixels in
a whole galaxy, then, can be seen to be somewhat over-simplistic, but
it serves as an important first step in evaluating the effectiveness of
the \pz technique, and greatly simplifies the interpretation of the results
of the remaining fitted parameters on a galaxy-by-galaxy basis.
Future applications of this technique will not be limited by
this assumption.

Each of the SEDs we use is simply a realization of the 4 dimensional
space of age, e-folding time, extinction and metallicity. To determine
the ``best fitting age'' of all the pixels in the galaxy, we use
equation~(\ref{eq:lkhood2}) to marginalize our likelihood estimator over
the age of the stellar population in each pixel. By multiplying the
likelihoods of all pixels for each of the 10 modeled ages, we determined
a unique value for the likelihood of the entire galaxy at any given
age. This likelihood is then maximized and the best fitting age is
determined. This assumption, while not strictly consistent with galaxy
formation scenarios, has the advantage of providing greater ease of
interpretation of the remaining parameters in the context of the
whole galaxy, in particular the star formation history of the system. We
stress here that we could have chosen not to make this assumption, but our
main concern at this stage of development of the \pz technique is to
understand what the best fitting template returned for each pixel is
able to tell us about the underlying physical conditions of the
galaxy. We will relax this assumption in a subsequent investigation.

Secondly, while simple to parametrize in the SEDs, the e-folding time of an
exponentially decaying star formation is not directly reconcilable with
the star formation rate per pixel, which would be a much more desirable
quantity to extract from the SEDs. To determine the star formation rate
per pixel in physical units, we use the empirical formula of \citet{ken98}
to link the UV luminosity of each pixel to its intrinsic star formation
rate:
\begin{equation}
\label{eq:sfr}
SFR \ (\Msun yr^{-1}) = 
1.4 \times 10^{-28} \ L_{\nu} \ \rm (ergs \ s^{-1} \ Hz^{-1}).
\end{equation}
We implicitly assume that this calibration holds at all redshifts.
Given the fact that the redshift of each pixel has been fixed to that
of its host galaxy, a pixel's absolute UV luminosity 
$L_{\nu}$ can be easily computed for any cosmological model, after
incorporating the
necessary K-correction. However, for our estimate to be reliable, we need
to assess how $L_{\nu}$ determined from aperture photometry for
the entire galaxy compares to that of all pixels added together. The
difference could be particularly noticeable for galaxies in which
a fixed aperture measures only their inner flux. This
comparison has the dual purpose of calculating the correction factor to
be applied to each pixel to reconcile the measurement of the UV
luminosities, and also to provide a scaling factor to be applied to all
pixels to convert the star formation rate returned by \citet{bru00} into
physical units. The spectrophotometric synthesis models of
\citet{bru00}, in fact, return a star formation rate normalized to a $1
\Msun$ galaxy. The results of this comparison are discussed below.

\subsection{Global vs Local Star Formation Rate}
\label{subsec:comparison}
We use the publicly available HDFN photometric catalog \citep{wil96}
together with either spectroscopic or photometric redshift estimates, to
determine each galaxy's absolute UV luminosity. We then make use of
equation~(\ref{eq:sfr}) to determine a galaxy's star formation
rate. Comparison with the star formation rate determined in a similar
fashion for all the pixels in a galaxy yields a normalization constant,
which in turn is used to calculate the star formation rate in $\Msun
yr^{-1}$ for individual pixels. As pointed out by our referee,
we could have avoided the computational steps required to calculate this
normalization constant by making direct use of the output of the Bruzual
and Charlot models. We updated our pipeline to make use of this direct
approach. To ensure consistency, we also ran a series of tests to
determine to what extent our global normalization would differ from that
obtained directly by the Bruzual and Charlot's models. 
Our tests revealed this discrepancy to be of the order of a few 
percent and suggest that for all practical purposes the two methods 
indeed are equivalent.

In Figure~\ref{fig:sfrs} the
top panel shows on the abscissa the 
star formation rate computed from aperture photometry for all galaxies
in the HDFN versus the star formation rate derived from the sum of all
pixels in the same galaxies. The line represents the one-to-one relation
and is not a fit to the data. The rms is $0.86$. 
The agreement is quite good, even though
the \pz result seems to slightly overestimate the overall SFR. This
behavior can be understood, as we mentioned earlier, by noticing that
for large galaxies a fixed aperture might not include all pixels.

The bottom panel shows the residual uncertainties in the measurement of
the star formation rate. Once again, aside from a few outliers, the
scatter is sufficiently small that we are confident our estimate of the
galaxy's UV luminosity obtained by adding the contribution of individual
pixels is indeed consistent with aperture photometry measurements. For
those galaxies with no UV detection, we have assumed an upper limit for
their fluxes as determined from the $\U$ rms maps. These objects are
generally small and, when using aperture photometry, the aperture is
likely to include many background pixels, which in turn give the
appearance of larger star formation rates. These outliers are clearly
visible in the residual plot.

The validity of the agreement displayed in Figure~\ref{fig:sfrs} will be
tested further when we calculate the star formation history of all the
galaxies in the HDFN and compare it to known results from the literature.

\subsection{Galaxy Maps}
\label{subsec:maps}
The result of the fit of the 2160 SED templates to a large spiral in the
HDFN is shown in Figure~\ref{fig:4gala}. The top left
map shows the galaxy in the F606W WFPC2 filter. 
The other three maps display the breakdown of the best fitting
template in each pixel according to values of color excess parametrized
in magnitudes, metallicity relative to the sun's, and star formation rate
in $\Msun/yr$. The redshifts of all pixels have been fixed to that of
the galaxy and the ``best fitting age'' of the galaxy has been computed as
described in \S~\ref{sec:analysis}.

This simple representation of the result of our
analysis allows for direct insight into the underlying characteristics of
the galaxy. In particular, in all three maps, the morphological
details of the galaxy are clearly recognizable. The dark knots
seen in the F606W image are also clearly visible in the other
maps. The SFR map distinctly displays the arm and inter-arm regions
with the former showing a SFR an order of magnitude higher than the latter.
A critical value that needs to be
associated with this estimate of the SFR per pixel, is the variance in
this estimate. As we will show later on, most of the sky pixels have
extremely large errors and thus, once the SFR is properly weighted, do
not contribute significantly to the overall SFR of the galaxy. These
pixels are mostly those at the periphery of the galaxy. This is evident
in Figure~\ref{fig:4gala}

The obscuration map is perhaps even more dramatic. While generally
displaying low color excess, large regions of the galaxy do show
considerable amounts of dust. The core of the galaxy, for example,
exhibits a $\sim 0.4$ magnitudes of obscuration along the line of sight. 
The arm regions seem to have pockets of lower obscuration surrounded by
regions at higher obscuration and generally appears to be quite patchy, in
accordance with observations \citep{tre98}. 
It is interesting to note that in the
obscuration map the knots present in the F606W image are even more
evident than in the SFR map where the SFR seems quite uniform across the
arm regions.

The metallicity map displays large regions of the galaxy at 
$Z \sim \frac{1}{50} Z_{\odot}$, but also small pockets in excess of
solar. The metallicity map seems to correlate quite well with the SFR
map in that all the regions with higher metallicity are indeed those
with ongoing star formation as one might expect. As in the SFR map, the
outskirts of the galaxy, where the contributions come from mostly sky
pixels, exhibit very low metallicities and obscuration, a sign that the
sky is generally described by templates with short e-folding times, no
extinction and low metallicity.

Thus, by making few simplifying assumptions, we are able to 
describe a galaxy in terms of its main physical processes at the
pixel-by-pixel level and hence discover a previously unseen view.
We can envision several scenarios in which this technique might
be employed to gain information on the local conditions within galaxies. In
particular, as far as the SFR of galaxies is concerned, \pz can
effectively separate high and low star forming regions in galaxies
and, provided a sufficiently large sample of galaxies is available,
trace their history. This information, together with a measure of the
fraction of old versus young star forming regions, will also provide
clues on the clustering properties of star formation. Coupled with
information on obscuration and metallicity, \pz should be able to return
at a minimum direct morphological information and, at best, a new
perspective on the evolution of galaxy morphology.

The galaxy in Figure~\ref{fig:4galb} display a substantially different
behavior. This galaxy is a face on spiral with clear star formation
activity in its spiral arm. The central regions of the galaxy exhibit
low star formation activity, solar metallicity and low obscuration. The
central region is surrounded by an extended star 
forming envelope which clearly traces the spiral design. Interestingly,
the lower right corner of the metallicity map, seems to suggest an
earlier burst of star formation in the history of this galaxy. This
region is characterized by solar and super-solar metallicity with low
obscuration and a moderate ongoing star formation rate.

Clearly, the use of the \pz technique can return particularly interesting
insights on the link between the morphology of a galaxy and it's
physical constituents. We will investigate this link in a forthcoming
paper. 

\subsection{Effect of photometric redshift uncertainty}
\label{subsec:zerrors}
Perhaps an even more important aspect of the \pz technique itself is the
understanding of the reliability of its estimate. While
Figure~\ref{fig:4gala} provides a great deal of immediate information on
the local characteristics of a galaxy, this information may at the same
time be misleading unless proved reliable.

To do so, we need to consider the source of our uncertainties. First and
foremost we need to address the effect the photometric redshift
error. To do so, we ran a series of Monte Carlo simulations on the
subsample of HDFN galaxies with known spectroscopic redshifts. We added
to the redshift of each of these galaxies a randomly distributed
Gaussian error with a dispersion ranging from $\sigma_z=0.05$ up to 
$\sigma_z=0.2$. We then ran the \pz pipeline on these galaxies and
recovered the difference in the best fitting template and hence in the
underlying physical properties described by the SEDs. 

The results are shown in Figure~\ref{fig:zeffect}. The method is rather
robust. Each panel represents the distribution of
deviations from the
initial estimate of the properties of galaxies embedded in the SED templates
in two photometric redshift error regimes. The solid line represents a
marginal error of $\sigma_z =0.05$ in redshift (an uncertainty typical
of the photometric redshift techniques) and produces an extremely narrow
peak around the correct value. The dashed line is the
result of a much broader redshift error distribution of $\sigma_z =0.2$.
Even with such a large uncertainty, the number of templates used by \pz
seem to be able to compensate for the errors. The distribution observed
in this case is almost unchanged.

\subsection{Error Maps}
\label{subsec:errors}
The other source of uncertainty in the \pz technique is the intrinsic 
error associated with the properties of each pixel derived from the best
fitting SED. To determine and understand the source and magnitude of
this error and how it propagates throughout the pipeline, we need to
make use of all the available information.

For each of the pixels in a galaxy, in fact, we have far more
information than the best fitting template. Each pixel is characterized
by a likelihood function result from the fit, i.e. the value of the
likelihood sampled along the 2160 templates. The maximum of this
likelihood returns the best fitting SED for a particular pixel, which
in turn can be decomposed into four quantities that uniquely determined the
SED, namely: age of the stellar population, e-folding time of an
exponentially decaying SFR, obscuration and metallicity.

Thus, the likelihood is, in reality, a complex four dimensional function
of the main physical quantities that drive the SED. The best fitting
template is nothing but the global maximum of this function. Here we are
interested in recovering a measure of our uncertainty on each of the
four axes of variability of the likelihood function. To do so, we
need to be able to ``collapse'' the four dimensional likelihood function
to each one of its dimensions in turn. This can be achieve by calculating the
marginalized likelihood function along each axis.

The results of this procedure for one of the
pixels in an HDFN galaxy is shown in Figure~\ref{fig:lkhoods}. The four
panels represent marginalized likelihood functions along different
axes. The shape of the four functions is quite typical of the behavior of
pixels in the HDFN galaxies and serves as a good illustrative example to
show how we calculate our uncertainties. While the four dimensional
likelihood was sampled at 2160 points corresponding to each of the
templates, the likelihoods shown in Figure~\ref{fig:lkhoods} are one
dimensional likelihood functions sampled at the values used to
parametrize the SED: age, e-folding time, color excess and
metallicity. Thus, the age likelihood function, for example, is sampled
at 10 different points corresponding to ages ranging from 0.1 to 15 Gyrs.
The other three panels are sampled at 6 different points corresponding
to the choices we made in \S~\ref{sec:temp} for obscuration,
metallicity and e-folding time.

For one dimensional likelihood functions with one degree of freedom,
such as those in Figure~\ref{fig:lkhoods}, a measure of the $1\sigma$
uncertainty (which corresponds to a $\Delta \chi^2 = 1$ or a normalized
likelihood of ${\cal L} = e^{-1/2}$) can be obtained simply by calculating the
range of parameter space intersected by the $1\sigma$ line. For this
particular pixel, for example, while the obscuration and metallicity
likelihood functions display a rather sharp peak, and thus produce small
uncertainties in these measures, age and e-folding time show much
broader likelihood functions which in turn correspond to larger
uncertainties. 

If we repeat the above exercise for all the pixels in all galaxies of
the HDFN, we can associate with each pixel four uncertainties. The results
are shown in Figure~\ref{fig:4err}, where the age dependence has been
suppressed as we compute the ``best fitting age'' for the galaxy using
the method described in \S~\ref{sec:analysis}.
As expected, those pixels with the
highest signal-to-noise ratio in all band-passes, typically pixels well
inside the object and commonly referred to as
``source pixels'', have relatively small errors in contrast with ``sky
pixels''. Direct comparison with the F606W image of
Figure~\ref{fig:4err} underlines how the SFR uncertainty, for example, is
indeed reliable only within the source and rapidly degrades outward.
Recalling the picture of the SFR we obtained from
Figure~\ref{fig:4gala}, we see that most of the high SFR at the outskirts
of the galaxy, once properly weighted by its uncertainty, carries a
small weight in the estimate of the total SFR in the galaxy. 
Perhaps even more dramatic is the case of the
metallicity uncertainty. The dark region represents pixels in the
galaxy where the metallicity estimate is reliable to less than
1 dex. These regions, are almost a one-to-one map of the optical image
shown. As we move outward the estimate degrades rapidly becoming
unreliable with uncertainties greater than 0.5-0.6 dex. A similar result
is seen in the color excess error map.

It now becomes clear that, without the aid of error maps such as the
ones in Figure~\ref{fig:4err}, any estimate of age, SFR, color excess and
metallicity are limited by the unknown reliability. 
Indeed, much of the computational time spent on the HDFN
galaxies by \pz was occupied by the calculation of the maximum
likelihood of the fit and its proper marginalization. By correctly
weighting all the pixels in a galaxy by the appropriate error maps, we
can move on to use the information contained in the galaxy maps of
Figure~\ref{fig:4gala} and measure
the star formation history (hereafter SFH) of galaxies in the HDFN.
Before doing so, however we will address the source of uncertainty and
degeneracies introduced by the particular SEDs selected for the 
analysis.

\subsection{SED Degeneracies}
\label{subsec:cex_vs_feh}
Finally, we would like to address the issue of template degeneracy. Due
to the large number of templates used, one might expect a certain degree
of degeneracy among the returned physical quantities that determine the
single SED of a pixel. To quantify this statement Figure~\ref{fig:met3D}
shows the time evolution of obscuration and metallicity. We have taken
the best fitting color excess and metallicity, as returned from the best
fitting template for each of the pixels in the HDFN galaxies, and fitted
a spline surface to examine the combined evolution. This approach
assumes this relation to be smooth on scales of $\Delta z \sim 0.5$,
which is the typical bin size for our star formation history
calculations (see the following Section).

At low redshift the extent of this degeneracy is clearly visible. Pixels
whose best fitting SED are characterized by solar and above solar
metallicities tend to inherently be more obscured than those at lower
metallicities. This behavior seems to be rather strong at redshifts below
one. As we look at pixels that belong to systems at higher redshift,
this behavior seems to change in favor of a flattening of the 
surface in the extinction direction suggesting that at
$z \sim 3$ obscuration is not a good indicator of the underlying
metallicity distribution. 

At the same time a general steady increase in
the metal content of pixels as a function of redshift is evident in
Figure~\ref{fig:met3D}. This behavior can perhaps be understood in
terms of a luminosity selection effect. 
It is well established that present-day galaxies
exhibit a clear trend between B-band luminosity and the oxygen abundance
of their H II regions \citep[see][]{ski89}.
This metallicity-luminosity relation
extends across morphological types and over 9 mag in luminosity and
appears to hold at least back to $z \sim 0.4$ \citep{kob99}. 
It seem plausible that at high redshift we are more likely 
(particularly in $\U$-band) to observe luminous star forming galaxies
which, in turn, are generally more metal rich \citep{nag00,mel02}. 
This behavior must be the result of the role
mass plays in determining the galaxy formation history, which sets the
chemical enrichment history. Mass must regulate either the rate at which
elements are produced by star formation or the ease with which they
can escape the gravitational potential of the galaxy (or both).

Degeneracies, particularly with a very large number of templates which
describe a wide range of underlying physical characteristics of the
stellar population, are to be expected. It is important, however, to be
aware of the interplay among the different physical parameters that
regulate the underlying SED in each pixel. To first order
Figure~\ref{fig:met3D} demonstrates that these degeneracies can induce
strong correlations among the physical quantities at play. The
information we gain by examining Figure~\ref{fig:met3D} can now be used
to interpret both the galaxy and error maps of
Figures~\ref{fig:4gala}, \ref{fig:4galb} and \ref{fig:4err}.

Finally, it is also important to underscore that the major source of 
uncertainty might indeed be the SED models themselves. In fact, some of 
the underlying spectral synthesis models we used might be inaccurate to
describe the properties of stellar populations or in the worst case
completely wrong. Spectra synthesis models for solar metallicity stellar
populations are now pretty well defined for optical photometry, but
such models are very much based on extrapolations for very sub-solar
and super-solar populations, and remain relatively untested in the UV.

\section{Star Formation History}
\label{sec:madau}
Using the framework of the previous two sections, we compute the SFR of
all pixels in all galaxies in the HDFN. At this stage of our analysis we
have information not only on the SFR of individual galaxies as a whole,
but, perhaps as importantly, on the SFR for each of the pixels in the
galaxy. As we shall see this information can be used to provide a
different view of the star formation history of the universe.

The galaxies in the HDFN span a range in luminosities and look-back
times. As a result, in order to be able to convert the star formation
histories of individual galaxies (and their pixels) into a comoving
average, we need to weight each galaxy (or pixel) by $1/V_{max}$ (Schmidt
1968; Bouwens et al. 1998). $V_{max}$ represents the maximum volume
within which a galaxy of a given apparent magnitude $m_{U}$ and
redshift $z_{gal}$ could still have been observed in the HDFN:
\begin{equation}
\label{eq:Vmax}
  V_{max} \equiv \int_\Omega \int^{z_{max}}_0
  \dfrac{\pd^2 V}{\pd \Omega \pd z}\, \pd z \pd \Omega 
\end{equation}
where $\Omega$ is the solid angle subtended by the HDFN, and $z_{max}$ is
the upper redshift limit of detectability for a galaxy with absolute
magnitude 
\beq
\label{eq:absmag}
M_{U} = m_{U} - 5 \log \dlz - 25 - K_{U}(z)\ , 
\eeq
at a luminosity distance $\dlz$ with a $\U$-band K-correction $K_{U}(z)$.

Since $z_{max}$ depends on the galaxy SEDs, we must account for the
$K$-correction when calculating $V_{max}$. The procedure adopted to
compute $V_{max}$ for both galaxies and pixels is identical, however,
the SED of each galaxy and of the pixels that belong to it, differ
considerably and so will their K-corrections. The photometric
redshift of each galaxy in the sample, $z_{gal}$ is derived by using the
maximum likelihood approach of equation~(\ref{eq:lkhood}), where the
small set of CWW templates is used to solve for both redshift and
SED. By contrast, for each of the pixels in the galaxy, the redshift
is held fixed at $z_{gal}$, while equation~(\ref{eq:lkhood2}) is maximized
for the best fitting SED selected from the 2160 templates we generate
using the Bruzual and Charlot spectrophotometric synthesis models
\citep{bru00}.  
For each of these pixels, the best fitting SED is used to
compute the appropriate redshift dependent K-correction in the
$\U$-band. This procedure requires equation~\ref{eq:absmag} to be
solved iteratively in order to solve for $z_{max}$.

Figure~\ref{fig:madau} shows the star formation history of galaxies and
their pixels in the HDFN. The crosses represent the individual galaxy
contribution to the star formation history of the universe as determined
by its $\U$-band luminosity converted into star formation rate according to
equation~(\ref{eq:sfr}). Below each galaxy, we show as
points the contribution to the star 
formation history of the universe of the individual pixels within each
galaxy. The comoving averaged contribution of all pixels in redshift intervals
is shown as filled circles. By comparison, we
show as open squares a measurements of the UV star formation history
taken from several sources in the literature.

We find the geometric mean of the star
formation rate in the HDFN to be $SFR \sim 0.19 \pm 0.22 {\ \rm
\Msun yr^{-1} Mpc^{-3}}$. The large dispersion is imputable mainly to a few
galaxies with SFR well above 10 ${\rm \Msun yr^{-1} Mpc^{-3}}$, while the bulk
of the population remains well below this value.

Immediately below each galaxy, we also show as points the contribution
to the star 
formation history of the universe of the individual pixels within each
galaxy. Typically the majority of the objects in the HDFN are rather
small, being just a few tens of pixels, but there are several large
galaxies containing a few thousand pixels. The mean SFR per pixel in the HDFN
$\left< SFR \right> \sim 0.004 \pm 0.052\ {\rm \Msun/yr/Mpc^3}$ is also
characterized by small star formation rates with a large dispersion
driven by the clumps of star formation present in large galaxies such as
the one shown in Figure~\ref{fig:4gala}. 

The comoving averaged contribution of all pixels in redshift intervals
is then calculated and is shown as filled circles. By comparison, we
show as open squares a measurements of the UV star formation history
taken from several sources in the literature \citep{gal95,lil95,mad96}. 
Our estimates
generally agree with the current picture of the star formation history
that has emerged in the last few years
\citep{lil95,cow96,ell96,gla95,yee96}: the SFR is, on 
average, a smooth function of redshift with a sharp increase up to $z
\sim 2$ and a decreasing trend at higher redshift. We will show in the
next section that this decline is directly related to the poor sampling
of the galaxy luminosity function at high redshifts. Once appropriate
luminosity functions are derived, the star formation history
is much flatter at $z>4$. This behavior is consistent with an exponential
decay of the SFR with time. In turn it also underscores that the bulk of
stars have formed at redshift higher than $z\sim 2$ \citep{mad98}.

For consistency, we
compare our results with other estimates not corrected for
obscuration. Despite the lack of an obscuration correction, the result
of using the luminosity functions below still supports a flat trend
beyond $z>2$.

\subsection{Luminosity Function Correction}
\label{subsec:LFcorr}
The galaxy luminosity function (LF) is one of the fundamental quantities
in observational cosmology. It provides us with a tool to investigate
the properties of the galaxy population as a whole. In the context of
this work, we are interested in determining how our knowledge, or lack
thereof, of the detailed form of the LF as a function of redshift, affects
the derived star formation history of the galaxies in the HDF. 

In particular, we want to be able to compensate for the fact that at high
redshifts we are sampling only the bright end of the galaxy luminosity
function and we are therefore systematically underestimating the
comoving density of star formation. As a result, somewhat different
absolute magnitude ranges are sampled at different redshifts and any
smooth functional form used to describe the LF will inevitably return
a redshift dependent description of the LF in its parameters. 

Nevertheless, a simple functional form, such as that introduced by
\citet{sch76} 
\begin{equation}
\label{eq:schLF}
\phi(M) = (0.4 \ln 10) \phi_{\star} \left[10^{0.4(M_{\star}-M)}\right]^{1+\alpha} \
                         \exp \left[-10^{0.4(M_{\star}-M)}\right] \ ,
\end{equation}
\noindent 
where $M_{\star}$ is the characteristic magnitude, $\alpha$ the 
faint-end slope and $\phi_{\star}$ the normalization factor,  
will allow an estimate of the total SFR to be
derived. We now use equation~(\ref{eq:schLF})
to estimate the LF within each of the redshift bins of
Figure~\ref{fig:madau}. 
We would like to emphasize here that we are not seeking to
trace the evolution of the Schechter parameters with
redshift. Instead, we can simply integrate the Schechter LF to a
magnitude fainter than our current $U_{300}$-band limiting magnitude, 
to derive a more complete estimate of the luminosity density than
available from the detected galaxies alone. 
Equation~(\ref{eq:sfr}) will then allow
us to convert this luminosity density to a measure of SFR density, as we
have done for both galaxies and pixels.

We compute the LF using the parametric maximum-likelihood method of
\citet[ hereafter STY, see also Lin et al. 1996]{san79}. 
The STY method is unbiased by density
inhomogeneities in the galaxy distribution and assumes a parametric
model for the galaxy LF. In our case, we take as our model for the
galaxy LF the Schechter function given in equation~(\ref{eq:schLF}) and
solve for $M_{\star}$ and $\alpha$. The STY method does not return
$\phi_{\star}$, the LF normalization constant. To calculate
$\phi_{\star}$, we used the minimum-variance estimator of
\citet{dav82}. Table~\ref{tab:LF} lists
the best fitting parameters of a Schechter LF in each redshift bin we
consider. At high redshift, the small number of (bright) galaxies
constrains the fit at the faint end of the luminosity
function. The value of $\alpha$ returned by our maximum
likelihood estimate there was unphysical, i.e. $\alpha<-2$. To
overcome this problem, the LF for galaxies at $z>3.5$ was determined by
assuming a conservative value of $\alpha=-1.1$.

The luminosity derived SFR density estimates are shown in
Figure~\ref{fig:madau} as filled squares and the 
last column of Table~\ref{tab:LF} contains the fraction of missing
comoving density of star formation. 
We have chosen a limiting magnitude of $U_{300,lim} = 28.5$ corresponding
to a $1\sigma$ detection threshold for point sources in the HDFN
\citep{con99}. At high redshift, the fraction of the total luminosity
in the detected galaxies decreases substantially compared to that at lower
redshifts. It appears evident that for $z<2$ we
miss almost half of the star formation and by $\sim 3$, we reach
90\%. Above this redshift, if our estimate of the luminosity function is
to be trusted, we miss virtually all the SFR.

The picture of the star formation rate that emerges from the
luminosity function derivation is that of an
exponentially increasing comoving density of star formation:
\begin{equation}
\label{eq:sfr_law} 
log\  \dot{\rho_*} \propto (1+z)^{0.12 \pm 0.02}
\end{equation}
This result is in agreement with the findings of \citet{lan02}. In
fact, current measurements miss a dominant fraction of the ultraviolet
luminosity density and hence of the star formation rate of the
universe. According to our results, the comoving density of star
formation rate determined from the UV luminosity density of sources in
the HDFN, increases monotonically with redshift out to at least redshift
of 5. This behavior can plausibly be explained either by a direct
increase of the UV luminosity density with redshift or by an increase in
the number of star forming regions as a function of redshift. 

To address this issue, we compute the fraction of the UV flux (and
hence of the star formation rate) needed to recover $90\%$ of the
detected star formation rate at each redshift bin in Figure~\ref{fig:madau}.
We divide the sample of HDFN galaxies (and their pixels) in two
broad sub-samples based on their rest frame (B-V) color: bluer and redder than
(B-V)=0.32 respectively.
This simple scheme is obviously not intended to be a rigorous separation of a
blue versus a red populations of galaxies (or pixels) in the HDFN. We are
interested in understanding the fraction of the total number of galaxies
and pixels needed to recover a given SFR as a function of
redshift. We argue that this quantity must be related to the history of
formation of the HDFN galaxies and to their current star formation
activity and perhaps linked to a galaxy's morphology.

Figure~\ref{fig:ceffect} shows the fraction of star formation rate
needed to recover $90\%$ of the total UV flux in the red and blue
population for galaxies (bottom panel) and pixels (top panel). The solid
line represents the red population, the dashed line the blue one.
The functional dependence with redshift has been smoothed on scales of
$z \sim 0.5$ in accordance with the typical width of the redshift bin in
Figure~\ref{fig:madau}. 

The red population (solid line) at the bottom of
Figure~\ref{fig:ceffect} represents the evolution of the fraction of
galaxies needed at any given redshift to recover 90\% of the SFR. For
$z>0.5$, this evolution seems to be rather similar to that of the blue
population (dashed line). This behavior suggests that up to relatively
high redshifts ($z \sim 3$) a comparable fraction of red and blue
galaxies are needed to give rise to the observed SFR. Below a
redshift of $z < 0.5$ the lack of a suitable blue population (indicated
by the size of the statistical error bar) suggests that there are simply
not enough low-redshift blue galaxies in the HDF to draw firm
conclusions. 

The top panel describes a quantitatively different behavior. The solid
and dashed lines represent the fraction of pixels in the red and blue
galaxies that accounts for $90\%$ of the SFR. At redshifts $z < 1$ we
need approximately $20\%$ of the pixels in red galaxies to account for
their SFR. As we move to higher redshifts, we observe a steady increase
in the number of pixels required to account for the UV flux out to $z
\sim 3$. At higher redshifts the number of galaxies is too small to draw
further  conclusions (see Figure~\ref{fig:madau}). This increase is a direct
consequence of the surface brightness dimming of the host galaxies that
harbor these pixels. This claim is supported by a simple exercise that
makes use of our ability to use the pixels' SED to coherently move
all the pixels in a galaxy to a new redshift taking into proper account the
surface brightness dimming due to the expansion of the universe. 

First we select a red, bright galaxy at low redshift. We then step in
redshift out to $z \sim 3$ in intervals of $\Delta z = 0.2$. At each
interval we use the SED of each pixel to compute the redshifted SED flux
convolved with the HST F300W U-passband filter. We then generate a
postage stamp image of the original galaxy and lay it on a grid of
pixels with the same resolution of our HDFN images
(i.e. $0.08\arcsec$/pixel). At each step we properly take into account
the new angular size of the galaxy by resampling the flux to match the
new size. Resampling is done using a 2D-spline fit to the flux at each
redshift. At each redshift, we also dim the flux to account for surface
brightness dimming. In this exercise we did not take into account the
intrinsic luminosity evolution of the galaxy. We were only interested in
measuring the effects of surface brightness dimming on the pixels.

This procedure returns the open squares in the top panel of
Figure~\ref{fig:ceffect}. The galaxy seems to follow the general trend
observed for the red pixels and hence its behavior is primarily dictated
by surface brightness effects. The typical statistical uncertainty
associated with each square is indicated in the corner of the figure.

The behavior of the blue galaxies is qualitatively different. The
fraction of blue pixels needed to recover $90\%$ of the flux sharply
increases to $\sim 30\%$ by redshift one and it remains almost unchanged
out to redshift $z \sim 3$. We can make use of our direct
knowledge of the SED in each pixel to artificially redshift a single
galaxy to high redshift and observe its behavior. The results are shown
in Figure~\ref{fig:ceffect} as open triangles. Surface brightness
dimming seems to be able to also explain the observed trend. It appears
that while the same fraction of galaxies are contributing to the SFR up
to $z\sim 3$, the number of pixels contributing to the flux differs. This
difference must be controlled by the relative distribution of pixels
{\it within} a galaxy as a function of redshift, and ultimately be
related to the morphological characteristics of galaxies.

At high redshift, mainly because of surface brightness dimming,
we need a similar fraction of blue and red pixels to recover most of the
star formation rate. By contrast, at low redshift the two populations of
pixels differ in their behavior suggesting that the \pz
technique can be used to probe this difference. In fact, the blue pixel
curve (top panel - dashed line), must hold some insight to the
intrinsic distribution of SFR in a galaxy. Where the star formation rate
is mostly concentrated at the center of a galaxy, as in the surface
brightness dimming induced case of high redshift galaxies, we need
roughly the same number of red and blue pixels to explain $90\%$ of the
SFR. However, a much smaller number of pixels is needed at lower redshift
to explain the red pixel behavior. This underscores the different
morphological characteristics of the host galaxies and suggests that we
are in the presence of more extended areas of star formation.

Ultimately, this last statement has to be closely connected with the
morphological characteristics of a galaxy. Clearly, the information
contained in the pixels and extracted using the \pz technique, can be
used as a probe of these morphological differences. We plan to address
the pixel-morphology connection in a forthcoming paper.

Finally, we can make use of our estimate of metallicity to compute the
comoving metallicity density of the universe. Our derived luminosity
densities were converted to metal enrichment rates to compare with the
predictions of \citet{pei95} and \citet{pei00} and with the $z > 2$ data
of \citet{mad96} (which, as noted, are based on 1500\AA\ luminosity
densities).  The conversion from L(2800\AA) to metal enrichment rate is
$2.2\times 10^{-23}$ M$_\odot$ yr$^{-1}$ W$^{-1}$ Hz. This value differs
from the that adopted by Madau et al.\ (1996) by a factor of
approximately 1.6 due, in part, to changes in the stellar synthesis
models of \citet{bru00}. Figure~\ref{fig:mer} shows
our results. Solid circles represent our direct estimate, while solid
squares represent our integrated luminosity function results.

Our uncorrected estimates are in good agreement with the
literature (Lilly et al, 1996 open triangle, Gallego et al., 1995 open
circles, Connolly et al., 1997 solid triangles) and with the theoretical
models of \citet{pei00} based on the evolution of the HI content in
damped Lyman-$\alpha$ systems. The luminosity function correction
shows a remarkable difference as a function of redshift from its
uncorrected values, in agreement the findings of this work and with
current measurements of the ultraviolet luminosity density \citep{lan02}.

\section{Conclusions and Future Applications}
\label{sec:end}
We analyze the photometric information contained in individual pixels of
galaxies in the Hubble Deep Field North (HDFN) using a new technique,
{\it pixel-z}, that combines predictions of evolutionary synthesis models
with photometric redshift template fitting. Each spectral energy
distribution  template is a result of modeling of the detailed physical
processes affecting gas properties and star formation efficiency. The
criteria chosen to generate the SED templates is that of sampling a
wide range of physical characteristics such as age, star formation rate,
obscuration and metallicity. By making use of a marginalized likelihood
error analysis were are also able to generate error maps that define
the reliability of the template fitting on pixel scales and allow for
the separation of the interplay among dust, metallicity and
star formation histories.

Our technique has the clear advantage of being able to probe the star
formation history of the universe independently of morphological
transformations. In fact, the curves shown in the top panel of
Figure~\ref{fig:ceffect} record the star formation activity of stars in
galaxies without any prior knowledge of their morphological distribution.
At the same time, however, \pz can be used to constrain the
distribution of star forming regions within galaxies in different
redshift ranges. This ``star formation clustering'' has
to be closely connected with the formation history and morphological
characteristics of a galaxy. Figure~\ref{fig:ceffect} strongly implies
that the \pz approach to studying the evolution of star formation in
galaxies returns a new description of the role SFR plays in determining
the morphological characteristics of system. This results suggests that
\pz does indeed provide us with new insights on galaxies.

Moreover, thanks to the large number of templates used and the range of
physical quantities they sample, \pz is able to return a detailed
snapshot of a system. Clear examples of this are shown in
Figures~\ref{fig:4gala} and \ref{fig:4galb} where two different galaxies
are examined. A key feature of \pz is its ability to make use of the
physical quantities recovered to examine the degeneracies in
the SED templates and the role they play in our understanding of
individual systems. Dust content, metallicity and star formation history
alter the colors of galaxies in ways which are by no means orthogonal
\cite[see]{kod99,dthm99}). Figure~\ref{fig:met3D}, for
example, underscored the importance of understanding the evolution of
the metallicity-obscuration relation in order to be able to interpret
the characteristics of an individual galaxies. Its important to mention
that these conclusions depend on the universal validity of the
Calzetti (1997) extinction law.

Finally, the integrated light from an entire galaxy or from individual
pixels represents an average over cosmic time of the stochastic star
formation episodes of individual galaxies, and will follow a relatively
simple dependence on redshift. By examining the evolution of the observed
UV flux, which in turn is proportional to SFR, we
hope to gain valuable insights on the mechanisms which may prevent the gas
within virialized dark matter halos from radiatively cooling and turning
into stars at early times, or on the epoch when galaxies exhausted their
reservoirs of cold gas \citep{mad96}.

However, it is apparent by inspection of Table~\ref{tab:LF} how
at high redshifts we are progressively sampling only the bright part of
the galaxy luminosity function. Hence, we compute an estimate of the possible
uncertainty in the global star formation rate induced by an incomplete
luminosity function by asking what portion of the total luminosity we
are missing. Figure~\ref{fig:madau} shows the luminosity function
corrected star formation rates and indicates that these corrections are
indeed quite strong. Our results show that the comoving density of star
formation rate, determined from the UV luminosity density of sources in
the HDFN, increases monotonically with redshift out to at least redshift
of 5. This behavior can plausibly be explained by a smooth increase of
the UV luminosity density with redshift coupled with an increase in the
number of star forming regions as a function of redshift. 
Furthermore, we find the overall metal enrichment rate history to be
consistent with the predictions of \citet{pei00} based on the
evolving HI content of Lyman-$\alpha$ QSO absorption line systems.

With the development of wider and more efficient space based cameras, such
as the ACS, deep, multi-band surveys will become the new testing grounds for
the \pz technique. At present \pz seems to be the best available tool to
help shed some light on the nature of star formation at moderate to
high redshift and its connection to galaxy formation processes.

\acknowledgments
We wish to thank Jeff Gardner for helpful discussions and Michael Fall
for providing his metal enrichment data in electronic form. 
We acknowledge the referee Roberto Abraham for comments and suggestions 
which helped to improve the overall clarity of the manuscript. 
AMH acknowledges support provided by the National Aeronautics and Space
Administration (NASA) through Hubble Fellowship grant
HST-HF-01140.01-A awarded by the Space Telescope Science Institute (STScI).
STScI is operated by the Association of Universities for Research in
Astronomy, Inc., under NASA contract NAS5-26555.
This work has been supported by NSF Career grant AST99-84924, by 
NASA Applied Information Systems Research grant NAG 5-9399 and by 
NASA LTSA Grant NAG5-8546.

{}

\clearpage

\begin{deluxetable}{cccccc}
\tablecolumns{5}
\tablewidth{0pc}
\footnotesize
\tablecaption{\scriptsize LUMINOSITY FUNCTION PARAMETERS\label{tab:LF}}
\tablehead{
\colhead{Redshift Range} & \colhead{$N_{gal}$} &
\colhead{$M_{\star}-5\log h^{a}$} & \colhead{$\alpha$} &
\colhead{$\phi_{\star} (h^3 Mpc^{-3})^{a}$} & \colhead{Missing Fraction$^{c}$}
}\startdata
$ 0.0 <z< 1.0$ & 355 & -17.71 $\pm$ 0.09 & -0.79 $\pm$ 0.12 & 0.082 $\pm$ 0.010 &  69\%\\
$ 1.0 <z< 2.0$ & 338 & -18.46 $\pm$ 0.09 & -0.95 $\pm$ 0.11 & 0.045 $\pm$ 0.002 &  33\%\\
$ 2.0 <z< 3.5$ & 174 & -19.40 $\pm$ 0.13 & -0.94 $\pm$ 0.15 & 0.035 $\pm$ 0.005 &  90\%\\
$ 3.5 <z< 5.0$ &  34 & -20.16 $\pm$ 0.29 & -1.10$^{b}$ $\pm$ 0.31 & 0.019 $\pm$ 0.004 & 98\%\\
\enddata
\tablenotetext{a}{$H_0 = 75\ h$ km s$^{-1}$ Mpc$^{-3}$. 
Errors are formal $1\sigma$ uncertainties.}
\tablenotetext{b}{Due to the small number of galaxies in this high
redshift bin, the faint end of the luminosity function was, at best,
poorly constrained. The value of $\alpha$ returned by our maximum
likelihood fit was unphysical, i.e. $\alpha<-2$. Hence, in this redshift
range alone, the value of $\alpha$ was held fixed.}
\tablenotetext{c}{Fraction of missing comoving density of star formation. 
In Figure~\ref{fig:madau} the observed points (filled circles) are affected by 
the incomplete sampling of the galaxy luminosity at faint magnitudes as 
a function of redshift. This in turn determines an underestimate of the 
SFR in each redshift interval. By integrating the
galaxy luminosity function in each redshift range out to 
$U_{300,lim} = 28.5$ and by transforming this luminosity into SFR
according to equation~(\ref{eq:sfr}), we are able to compute
the necessary correction to the points in Figure~\ref{fig:madau}
(filled squares).}
\end{deluxetable}

\clearpage

\begin{figure}
\caption{Distribution of pixels as a function of redshift in a bright,
large spiral galaxy in the HDFN. For each pixel a photometric redshift is
computed using the template fitting technique. For each of the pixels,
the resulting photometric redshift estimate, is simply a function of the
flux in each of the available band-passes. When applied to the entire
galaxy (i.e. the sum of the contributions from individual pixels), the
photometric redshift returned should correlate with the number of pixels
that have photometric redshifts near the galaxy overall redshift.
\label{fig:phist1}}
\end{figure}

\begin{figure}
\caption{Same as Figure~\ref{fig:phist1}, but for a small and faint
galaxy in the HDFN. The more interesting features of this figure
reside in the number and strength of the 3 peaks shown and in the 
information they might contain. 
One can envision a scenario in which two objects at different
redshifts, but close in angular distance, might be separated by using
this method. This appears to be on such case \citep{con99}. See the text
for details. 
\label{fig:phist2}}
\end{figure}

\begin{figure}
\caption{Top: Star formation rate computed from aperture photometry for
all galaxies in the HDFN versus the star formation rate derived from the
sum of all pixels in the same galaxies. The line represents the
one-to-one relation and is not a fit to the data. The agreement is quite
good, even though the \pz result seems to slightly overestimate the
overall SFR. This behavior can be understood by noticing that for large
galaxies a fixed aperture might not include all pixels.  Bottom:
Residual uncertainties in the measurement of the star formation
rate. For those galaxies with no UV detection, we have assumed an upper
limit for their fluxes as determined from the $\U$ rms maps.  These
objects are generally small and, when using aperture photometry, the
aperture is likely to include many background pixels, which in turn give
the  appearance of larger star formation rates. These outliers are
clearly visible in the residual plot.\label{fig:sfrs}}
\end{figure}

\begin{figure*}
\caption{Result of the fit of the 2160 SED templates to a large spiral
galaxy in the HDFN. The top left map shows the galaxy in the F606W WFPC2
filter.  The other three maps display the breakdown of the best fitting
template in each pixel according to values of color excess parametrized
in magnitudes, metallicity relative to the sun's and star formation rate
in $\Msun/yr$. The redshifts of all pixels have been fixed to that of
the galaxy and the ``best fitting age'' of the galaxy has been computed
as described in \S~\ref{sec:analysis}. See the text for details.
\label{fig:4gala}}
\end{figure*}

\begin{figure}
\caption{Same as Figure~\ref{fig:4gala} for another galaxy in the HDFN.
\label{fig:4galb}}
\end{figure}

\begin{figure}
\caption{Estimate of the photometric redshift error propagation in the
\pz pipeline. Each panel represents the distribution of
deviations from the
initial estimate of the properties of galaxies embedded in the SED templates
in two photometric redshift error regimes. The solid line represents a
marginal error of $\sigma_z =0.05$ in redshift (an uncertainty typical
of the photometric redshift techniques) and produces an extremely narrow
peak around the correct value. The dashed line is the
result of a much broader redshift error distribution of $\sigma_z =0.2$. Not
surprisingly the distribution of deviations broadens, but still within
very acceptable limits. 
\label{fig:zeffect}}
\end{figure}

\begin{figure}
\caption{Marginalized likelihood functions along different
axes. While the four dimensional
likelihood was sampled at 2160 points corresponding to each of the
templates, the likelihoods shown are one
dimensional likelihood functions sampled at the values used to
parametrize the SED: age, e-folding time, color excess and
metallicity. Thus, the age likelihood function, for example, is sampled
at 10 different points corresponding to ages ranging from 0.1 to 15 Gyrs.
The other three panels are sampled at 6 different points corresponding
to the choices we made in \S~\ref{sec:temp} for obscuration,
metallicity and e-folding time. The horizontal line represents a
$1\sigma$ uncertainty. The likelihoods shown are the result of
a polynomial fit to the data. 
\label{fig:lkhoods}}
\end{figure}

\begin{figure}
\caption{Error maps for the galaxy shown in
Figure~\ref{fig:4gala}. These maps are obtained by computing the
marginalized likelihood for each pixel in the galaxy as described in
\S~\ref{subsec:errors}. Note that the SFR is
indeed reliable only within the source and rapidly degrades outward.
Most of the high SFR at the outskirts of the galaxy, once properly
weighted by its uncertainty, carries a small weight in the estimate of
the total SFR. 
\label{fig:4err}}
\end{figure}

\begin{figure}
\caption{Evolution of obscuration and metallicity as a function of
redshift for all pixels in the HDF-N. At low
redshift pixels whose best fitting SED are characterized by solar and
above solar metallicities tend to inherently be more obscured than those
at lower metallicities. The nature of this degeneracy is rather strong
at redshifts below one. At higher redshifts,
this behavior changes in favor of a flattening of the 
surface in the extinction direction suggesting that at
$z \sim 3$ obscuration is not a good indicator of the underlying
metallicity distribution. A general steady increase in
the metal content of pixels as a function of redshift is also evident. This
behavior can be understood in terms of a luminosity selection
effect whereby higher luminosity systems are also preferentially metal
rich.\label{fig:met3D}}
\end{figure}

\begin{figure}
\caption{Star formation history of galaxies and their pixels in the HDFN.
The crosses represent the individual galaxy
contribution to the star formation history of the universe as determined
by its $\U$-band luminosity. Below each galaxy, we show as
points the contribution to the star 
formation history of the universe of the individual pixels within each
galaxy. The comoving averaged contribution of all pixels in redshift intervals
is shown as filled circles. By comparison, we
show as open squares a measurements of the UV star formation history
taken from several sources in the literature. The luminosity derived SFR
density estimates are show as filled squares and the  
last column of Table~\ref{tab:LF} contains the fraction of missing
comoving density of star formation.
\label{fig:madau}}
\end{figure}

\begin{figure}
\caption{Fraction of the UV flux (and
hence of the star formation rate) needed to recover $90\%$ of the star
formation rate in each redshift bin. A blue and a red sample 
were extracted from the HDFN galaxies based on B-V rest frame colors.
The top panel shows the fraction needed to recover $90\%$ of the total
UV flux for HDFN galaxies. The bottom panel looks at pixels. The solid
line represents the red population, the dashed line the blue one.
The functional dependence with redshift has been smoothed on scales of
$z \sim 1$ in accordance with the typical width of the redshift bin in
Figure~\ref{fig:madau}.
\label{fig:ceffect}}
\end{figure}

\begin{figure}
\caption{Metal enrichment of the universe rate as a function of
redshift. Solid circle
represent our direct estimate, while solid 
squares represent our luminosity function correction.
Our uncorrected estimates are in good agreement with the
literature (Lilly et al., 1996 open triangle, Gallego et al., 1995 open
circles, Connolly et al., 1997 solid triangles) 
and with the theoretical models of \citet{pei00} (solid and dashed
lines). The luminosity function correction
shows a remarkable difference as a function of redshift from its
uncorrected values. See the text for details.
\label{fig:mer}}
\end{figure}

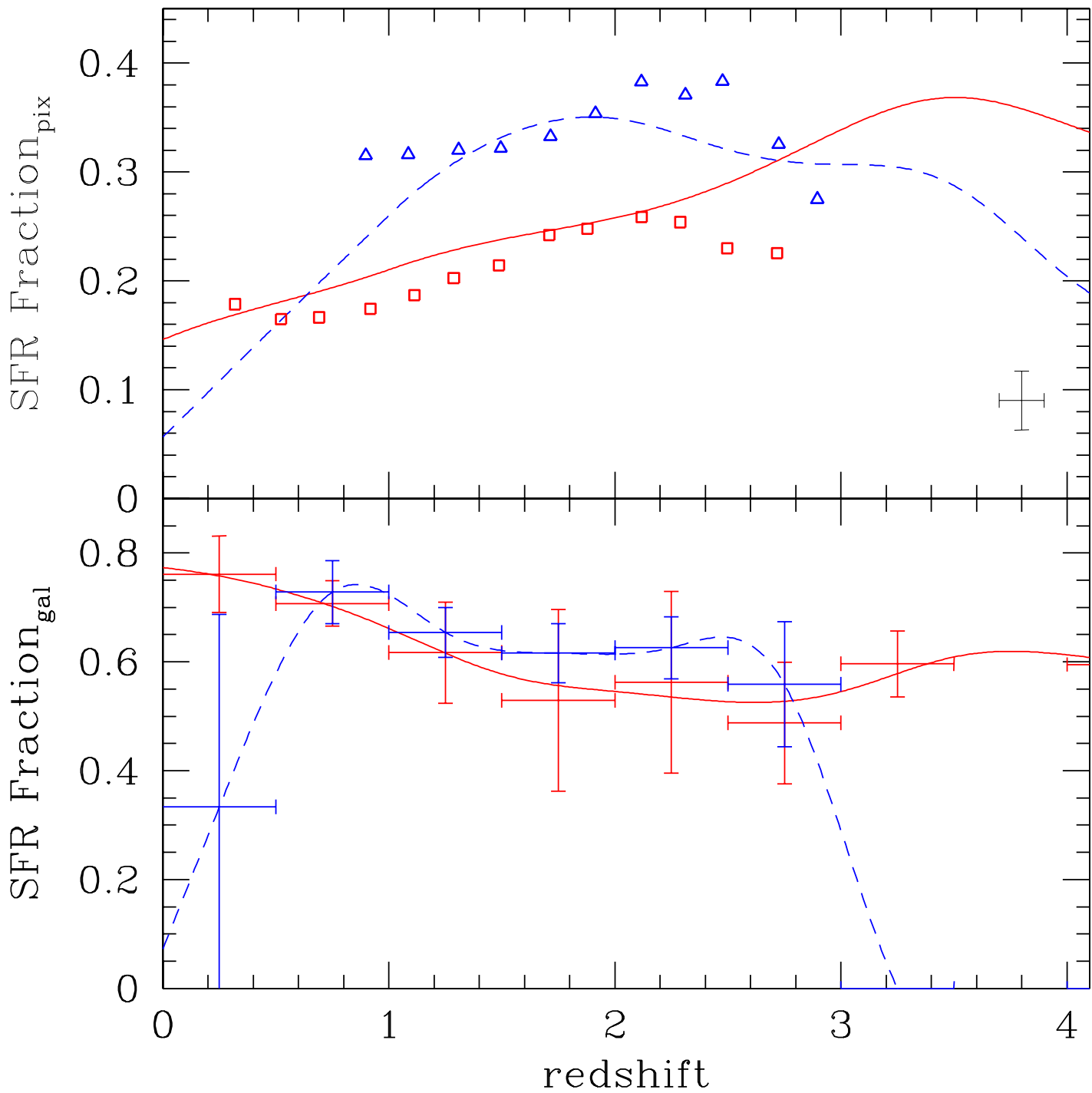
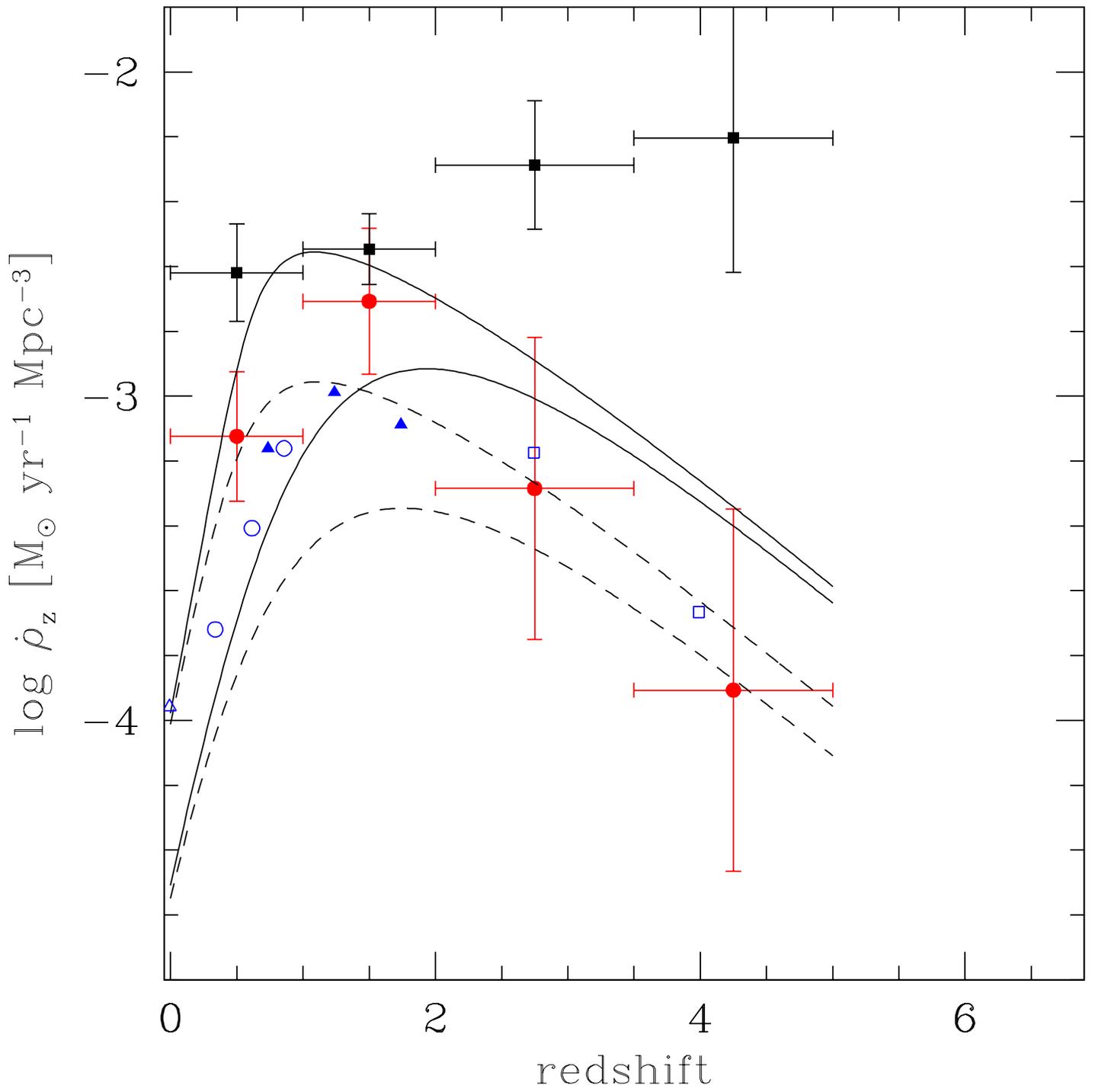
\end{document}